\documentclass[graybox]{svmult}

\usepackage{graphicx}

\usepackage{amsmath}
\usepackage{color}
\usepackage{float}
\usepackage[rflt]{floatflt}

\usepackage[latin1]{inputenc}
\usepackage{amsfonts}
\usepackage{tikz} 
\usetikzlibrary{matrix}
\usepackage[english]{babel}
\usepackage[latin1]{inputenc}
\usepackage[T1]{fontenc}
\usepackage{ae}

\usepackage{url}

\allowdisplaybreaks[4]

\usepackage{color}
\usepackage{tikz}
\usetikzlibrary{matrix}
\usepackage{cite}           
\usepackage{mathptmx}       
\usepackage{helvet}         
\usepackage{courier}        
\usepackage{type1cm}        
\usepackage{makeidx}         
\usepackage{graphicx}        
\usepackage{multicol}        
\usepackage[bottom]{footmisc}
\newcommand{\ELI}{{\rm ELi}}

\usepackage{tikz}
\usepackage{etex, xy}
\xyoption{all}
\usetikzlibrary{matrix}

\definecolor{blaugrau}{rgb}{0.796887, 0.789075, 0.871107}

\newcounter{linectr}

\allowdisplaybreaks[4]

\newcounter{mmacnt}
\def\restartmma{\setcounter{mmacnt}{0}}
\restartmma \catcode`|=\active
\def|#1|{\mathrm{#1}}
\catcode`|=12
\newenvironment{mma}{
 \par
 \catcode`|=\active
 \parskip=2pt\parindent=0pt 
 \small
 \def\In##1\\{%
   \def\linebreak{\hfill\break\null\qquad}%
   \refstepcounter{mmacnt}
   \hangindent=2.5em\hangafter=0
   \leavevmode
   \llap{\tiny\sffamily In[\arabic{mmacnt}]:=\kern.5em}%
   \mathversion{bold}\scriptsize$\tt\bf\displaystyle##1$\normalsize
   \mathversion{normal}\par
 }%
 \def\Print##1\\{%
   \def\linebreak{\hfill\break}%
   \hangindent=2.5em\hangafter=0
   \leavevmode\scriptsize ##1\par}%
 \def\Out##1\\{%
   \vspace*{-0.2cm}\def\linebreak{$\hfill\break\null\hfill$}%
   \kern\abovedisplayskip\par
   \hangindent=2.5em\hangafter=0
   \leavevmode
   \llap{\tiny\sffamily Out[\arabic{mmacnt}]=\kern.5em}
   \scriptsize$\displaystyle\tt##1$\normalsize\hfill\null\par
   \kern\belowdisplayskip\vspace*{-0.3cm}
 }%
 \def\Warning##1##2\\{%
   \def\linebreak{\hfill\break}%
   \hangindent=2.5em\hangafter=0
   \leavevmode
   {\scriptsize##1 : ##2}\par}%
}{%
 \par\smallskip
}

\spnewtheorem{psdefinition}{Definition}[section]{\bf}{}
\spnewtheorem{pstheorem}[psdefinition]{Theorem}{\bf}{}
\spnewtheorem{psproposition}[psdefinition]{Proposition}{\bf}{}
\spnewtheorem{pscorollary}[psdefinition]{Corollary}{\bf}{}
\spnewtheorem{pslemma}[psdefinition]{Lemma}{\bf}{}
\spnewtheorem{psremark}[psdefinition]{Remark}{\bf}{}
\spnewtheorem{psexample}[psdefinition]{Example}{\bf}{}
\spnewtheorem{psconvention}[psdefinition]{Convention}{\bf}{}

\newcommand{\modd}{{\rm mod}}

\newcommand{\Li}{{\rm Li}}
\newcommand{\Mvec}{{\rm \bf M}}

\usepackage{rotating}

\usepackage{graphicx}

\usepackage{color}

\newenvironment{fshaded}{%
\MakeFramed {\FrameRestore}
}%
{\endMakeFramed}

\usepackage{tikz}
\usetikzlibrary{matrix}

\allowdisplaybreaks[4]

\newcommand*\pFqskip{8mu}
\catcode`,\active
\newcommand*\pFq{\begingroup
        \catcode`\,\active
        \def ,{\mskip\pFqskip\relax}%
        \dopFq
}
\catcode`\,12
\def\dopFq#1#2#3#4#5{%
        {}_{#1}F_{#2}\biggl[\genfrac..{0pt}{}{#3}{#4};#5\biggr]%
        \endgroup
}

\makeindex             

\begin{document}

\title*{{\footnotesize  \sf DESY 18-143,~DO-TH 17/08}\\
Iterative Non-iterative Integrals in Quantum Field Theory}
\titlerunning{Iterative Non-iterative Integrals in Quantum Field Theory}
\author{Johannes Bl\"umlein}
\institute{Johannes Bl\"umlein, Deutsches Elektronen-Synchrotron, DESY, Platanenallee 6, D-15738 
Zeuthen, Germany, \email{Johannes.Bluemlein@desy.de}}
%
%
\maketitle

\abstract{Single scale Feynman integrals in quantum field theories obey difference or differential
equations with respect to their discrete parameter $N$ or continuous parameter $x$. The analysis
of these equations reveals to which order they factorize, which can be different in both cases.
The simplest systems are the ones which factorize to first order. For them complete solution 
algorithms exist. The next interesting level is formed by those cases in which also irreducible 
second order systems emerge. We give a survey on the latter case. The solutions can be obtained
as general $_2F_1$ solutions. The corresponding solutions of the associated inhomogeneous differential 
equations form so-called iterative non-iterative integrals. There are known conditions under which one may 
represent the solutions by complete elliptic integrals. In this case one may find representations in terms 
of meromorphic modular functions, out of which special cases allow representations in the framework of 
elliptic polylogarithms with generalized parameters. These are in general weighted by a power of 
$1/\eta(\tau)$, where $\eta(\tau)$ is Dedekind's $\eta$-function. Single scale elliptic solutions emerge in 
the $\rho$-parameter, which we use as an illustrative example. They also occur in the 3-loop QCD 
corrections to massive operator matrix elements and the massive 3-loop form factors. 
}

\section{Introduction}
\label{sec:1}

\vspace*{1mm}
\noindent
In this paper a survey is presented on the classes of special functions, represented by 
particular integrals, to which presently known single scale Feynman-integrals evaluate.
Zero-scale integrals, also playing an important role in elementary particle physics,
are given by special numbers, see e.g. 
\cite{Blumlein:2009cf,Ablinger:2011te,Ablinger:2013cf,Ablinger:2014bra,Laporta:2017okg}.
To this class the expansion coefficients of the $\beta$-functions 
\cite{Baikov:2016tgj,Herzog:2017ohr,Luthe:2017ttg} and the renormalized masses, 
as well as $(g-2)$ \cite{Laporta:2017okg}, do belong. Single scale quantities depend on one 
additional parameter as e.g. the Mellin variable $N$, a momentum fraction or scale-ratio 
$x \in [0,1]$ and similar quantities. To this class contribute e.g. the massless Wilson coefficients 
\cite{Vermaseren:2005qc}, the anomalous dimensions \cite{Moch:2004pa,Vogt:2004mw,Ablinger:2017tan},
and the massive Wilson coefficients at large virtualities $Q^2$ 
\cite{Ablinger:2010ty,Ablinger:2014vwa,Ablinger:2014nga,Ablinger:2014lka,Behring:2014eya}.   

It is now interesting to see which function spaces span the analytic results of these quantities.
Traditionally two representations are studied:~{\it i)} the Mellin space representation following
directly from the light cone expansion  \cite{LCE} and ~{\it ii)} its Mellin inversion, the 
$x$-space representation, with $x$ the Bjorken variable or another ratio of invariants, which in 
particular has phenomenological importance.

In the first case the quantities considered obey difference equations, while in the second case
the corresponding equations are differential equations which are related to the former ones 
\cite{NOERLUND}. In all the cases quoted above either the recurrences or the differential operators
or both factorize at {\it first order} after an appropriate application of decoupling formalisms
\cite{Zuercher:94,ORESYS,Bluemlein:2014qka}. Due to this all these cases can be solved 
algorithmically in any basis of representation, as has been shown in Ref.~\cite{Ablinger:2015tua}.
In $N$-space the solution is then possible using C. Schneider's packages {\tt Sigma} 
\cite{SIG1,SIG2}, {\tt EvaluateMultiSum} and {\tt SumProduction} \cite{EMSSP}. Corresponding solutions
in $x$-space can be obtained using the method of differential equations \cite{DEQ,Ablinger:2015tua}.
This applies both to the direct calculation of the Feynman diagrams as well as to the calculation of 
their master integrals which are obtained using the integration by parts relations \cite{IBP}.

The above class of problems is the first one in a row. In general, the difference and differential equation 
systems do not decouple at first order, but will have higher order subsystems, i.e. of second, third, fourth 
order etc., cf.~\cite{Blumlein:2018cms}. Since the first order case is solved completely 
\cite{Ablinger:2015tua}, it is interesting to see which mathematical spaces represent the solution.
In $N$-space next to pure rational function representations the nested harmonic sums emerge 
\cite{Vermaseren:1998uu,Blumlein:1998if}. They correspond to the harmonic polylogarithms in $x$-space 
\cite{Remiddi:1999ew}. At the next level generalized harmonic sums and iterated
integrals of the Kummer-Poincar\'e type appear  \cite{KUMMER1,Moch:2001zr,Ablinger:2013cf}. These are followed 
by iterated integrals over cyclotomic letters \cite{Ablinger:2011te} and further by square-root valued letters, 
cf.~\cite{Ablinger:2014bra} and their associated sums and special constants, cf. also 
\cite{Ablinger:2013eba,Ablinger:2013jta,Blumlein:2018cms}. This chain of functions is probably not complete yet, as 
one might think of more general Volterra-iterated integrals and their associated nested sums, which are also obeying 
first order factorization. The main properties of these functions, such as their shuffling relations 
\cite{SHUF,Blumlein:2003gb} and certain general transformations are known. Most of the corresponding mathematical 
properties to effectively handle these special functions are implemented in the package {\tt HarmonicSums} 
\cite{HARMONICSUMS,Ablinger:PhDThesis,Ablinger:2011te,Ablinger:2013cf,Ablinger:2014bra}.

The next important problem is, how to deal with cases in which neither recurrences in $N$-space nor differential 
equations in $x$-space factorize at first order. Here, the general solution can be given by so-called
iterative non-iterative integrals\footnote{Iterative non-iterative integrals have been
introduced by the author in a talk on the 5th International Congress on Mathematical Software, held at
FU Berlin, July 11-14, 2016, with a series of colleagues present, cf. \cite{ICMS16}.}, implied by the representation 
of the solution through the variation of constant \cite{EULLAG} at {\it any order} of non-decoupling. 
This, of course, is a quite general statement, calling for refinement w.r.t. the corresponding special functions at 
non-decoupling to 2nd, 3rd, etc. order. In this article we will deal with the 2nd order case, discussing results, 
which have been obtained in Refs.~\cite{Ablinger:2017bjx,Blumlein:2018aeq} and by other authors recently.
At present, in the singly variate case, the highest order of non-decoupling being observed 
is 2nd order, see e.g. 
Refs.~\cite{SABRY,TLSR1,Caffo:2002ch,Laporta:2004rb,TLSR2,TLSR2a,Bailey:2008ib,Broadhurst:2008mx,
TLSR3,BLOCH2,TLSR4,Adams:2015gva,Adams:2015ydq,TLSR5,Adams:2014vja,Remiddi:2016gno,Adams:2016xah,
TSLR6,Passarino:2016zcd,Ablinger:2017bjx,Blumlein:2018aeq,vonManteuffel:2017hms,Adams:2017ejb,Bogner:2017vim}.

\section{Second order differential equations and \boldmath $_2F_1$ Solutions}
\label{sec:2}

\vspace*{1mm} \noindent
We consider the non-factorizable problem of order two in $x$-space. It is given by a corresponding differential 
equation of second order, usually with more than three singularities. Below we will give illustrations for equations
which emerge in the calculation of the $\rho$-parameter \cite{Grigo:2012ji,Blumlein:2018aeq}. These are Heun
differential equations \cite{HEUN}. A second order differential equation with three singularities can be mapped into 
a Gau\ss{}'
differential equation \cite{GAUSS}. In case of more singularities, this is possible too, however, the argument
if the $_2F_1$ function is a rational function through which the other singularities are described. It is of 
advantage to look for the latter type solutions, since the properties of the $_2F_1$ function are very well known
\cite{HYP,SLATER,KUMMER,RIEMANN,GOURSAT}.

We consider the non-factorizable linear differential equations of second order
\begin{eqnarray}
\left[\frac{d^2}{dx^2} + p(x) \frac{d}{dx} + q(x)\right] \psi(x) = N(x)~, 
\label{eq:D2}
\end{eqnarray}
with rational functions $r(x) = p(x), q(x)$, which may be decomposed into\footnote{In 
the present case only single poles appear; for Fuchsian differential equations $q(x)$ 
may have double poles.}
\begin{eqnarray}
r(x) = \sum_{k=1}^{n_r} \frac{b_k^{(r)}}{x - a_k^{(r)}},~~~~~a_k^{(r)},b_k^{(r)} \in \mathbb{Z}~.
\label{eq:rx}
\end{eqnarray}
The homogeneous equation is solved by the functions $\psi_{1,2}^{(0)}(x)$, which are linearly 
independent, i.e. their Wronskian $W$ obeys
\begin{eqnarray}
W(x) = \psi_1^{(0)}(x) \frac{d}{dx} \psi_2^{(0)}(x)
     - \psi_2^{(0)}(x) \frac{d}{dx} \psi_1^{(0)}(x) \neq 0~.
\end{eqnarray}
The homogeneous Eq.~(\ref{eq:D2}) determines the well-known differential equation for $W(x)$
\begin{eqnarray}
\frac{d}{dx} W(x) = -p(x) W(x)~,
\end{eqnarray}
which, by virtue of (\ref{eq:rx}), has the solution
\begin{eqnarray}
\label{eq:WR}
W(x) = \prod_{k=1}^{n_p} \left(\frac{1}{x-a_k^{(p)}}\right)^{b_k^{(p)}}~,
\end{eqnarray}
normalizing the functions $\psi_{1,2}^{(0)}$ accordingly.
A particular solution of the inhomogeneous equation (\ref{eq:D2}) is then obtained by Euler-Lagrange 
variation of constants \cite{EULLAG}  
\begin{eqnarray}
\label{eq:INHOM}
\hspace*{-3mm}
\psi(x) &=&~~\psi_1^{(0)}(x) \left[C_1 - \int dx~\psi_2^{(0)}(x) n(x)\right] 
 + \psi_2^{(0)}(x) \left[C_2 + \int dx~\psi_1^{(0)}(x) n(x) \right],
\label{eq:DI}
\end{eqnarray}
with 
\begin{eqnarray}
n(x) = \frac{N(x)}{W(x)} 
\end{eqnarray}
and two constants $C_{1,2}$ to be determined by special physical requirements. 
As examples we consider the systems of differential equations given in \cite{Grigo:2012ji} for the $O(\varepsilon^0)$ 
terms in the dimensional parameter. These are master integrals determining the $\rho$-parameter at general fermion 
mass ratio at 3-loop order. The corresponding equations read
\begin{eqnarray}
\label{eq:one}
0 &=& 
\frac{d^2}{dx^2} f_{8a}(x) 
+\frac{9-30 x^2+5 x^4}{x(x^2-1)(9-x^2)} \frac{d}{dx} f_{8a}(x)
-\frac{8 (-3+x^2)}{(9-x^2)(x^2-1)} f_{8a}(x)
\nonumber\\ &&
-\frac{32 x^2}{(9-x^2)(x^2-1)} \ln^3(x)
+\frac{12 (-9+13 x^2+2 x^4)}{(9-x^2)(x^2-1)} \ln^2(x)
\nonumber\\ &&
-\frac{6 (-54+62 x^2+x^4+x^6)}{(9-x^2)(x^2-1)} \ln(x)
+\frac{-1161+251 x^2+61 x^4+9 x^6}{2 (9-x^2)(x^2-1)}\\
\label{eq:f9bb}
f_{9a}(x) &=& -\frac{5}{8}(-13 - 16 x^2 + x^4) 
+  \frac{x^2}{2} (-24 + x^2) \ln(x)  + 3 x^2 \ln^2(x) - \frac{2}{3} f_{8a}(x)
\nonumber\\ &&
+  \frac{x}{6} \frac{d}{dx} f_{8a}(x).
\end{eqnarray}
There are more equations contributing to the problem, cf.~\cite{Ablinger:2017bjx}, in which in the inhomogeneity
more harmonic polylogarithms $H_{\vec{a}}(x)$ \cite{Remiddi:1999ew} contribute. Eq.~(\ref{eq:one}) is an Heun equation
in $x^2$. Its homogeneous solutions, \cite{Ablinger:2017bjx}, are:
\begin{eqnarray}
\label{eq:ps1a}
\psi_{1a}^{(0)}(x) &=& \sqrt{2 \sqrt{3} \pi} 
\frac{x^2 (x^2-1)^2 (x^2-9)^2}{(x^2+3)^4}
\pFq{2}{1}{{\tfrac{4}{3}},\tfrac{5}{3}}{2}{z}
\\
\label{eq:ps2a}
\psi_{2a}^{(0)}(x) &=& \sqrt{2 \sqrt{3} \pi}
\frac{x^2 (x^2-1)^2 (x^2-9)^2}{(x^2+3)^4}
\pFq{2}{1}{{\tfrac{4}{3}},\tfrac{5}{3}}{2}{1-z},
\end{eqnarray}
with
\begin{eqnarray}
z = z(x) = \frac{x^2(x^2-9)^2}{(x^2+3)^3}~.
\end{eqnarray}
The Wronskian for this system is
\begin{eqnarray}
\label{eq:W1}
W(x) = x (9 - x^2) (x^2-1).
\end{eqnarray}
These are single-$_2F_1$ solutions, however, they are not given by single elliptic integrals.
One first uses contiguous relations and then mappings according to the triangle group 
\cite{TAKEUCHI,IVH,VANH1} and the algorithm described in appendix A of \cite{Ablinger:2017bjx} to 
obtain the solutions
\begin{eqnarray}
\label{eq:ps1b}
\psi_{1b}^{(0)}(x) &=& \frac{\sqrt{\pi}}{4 \sqrt{6}} \Biggl\{
- (x-1)(x-3)(x+3)^2 \sqrt{\frac{x+1}{9-3x}} 
\pFq{2}{1}{{\tfrac{1}{2}},\tfrac{1}{2}}{1}{z}
\nonumber\\
&& + (x^2+3)(x-3)^2     \sqrt{\frac{x+1}{9-3x}}
\pFq{2}{1}{{\tfrac{1}{2}},-\tfrac{1}{2}}{1}{z} \Biggr\}
\\
\label{eq:ps2b}
\psi_{2b}^{(0)}(x) &=&
\frac{2\sqrt{\pi}}{\sqrt{6}}\Biggl\{x^2 \sqrt{(x+1)(9-3x)}
\pFq{2}{1}{{\tfrac{1}{2}},\tfrac{1}{2}}{1}{1-z}
\nonumber\\ && 
+\frac{1}{8} \sqrt{(x+1)(9-3x)} (x-3)(x^2+3)
\pFq{2}{1}{{\tfrac{1}{2}},-\tfrac{1}{2}}{1}{1-z}\Biggr\},
\end{eqnarray}
where
\begin{eqnarray}
\label{eq:z1}
z(x) = -\frac{16x^3}{(x+1)(x-3)^3}~.
\end{eqnarray}
and 
\begin{eqnarray}
\label{ell:K}
\pFq{2}{1}{{\tfrac{1}{2}},\tfrac{1}{2}}{1}{z}  &=& \frac{2}{\pi} {\bf K}(z) \\
\label{ell:E}
\pFq{2}{1}{{\tfrac{1}{2}},-\tfrac{1}{2}}{1}{z} &=& \frac{2}{\pi} {\bf E}(z)~,
\end{eqnarray}
cf.~\cite{TRICOMI}. Here ${\bf K}$ denotes the elliptic integral of the first and ${\bf 
E}$
the elliptic integral of the second kind.

Analyzing the criteria given in \cite{RHEUN1,RHEUN2} one finds, that the solution (\ref{eq:ps1b},\ref{eq:ps2b}) 
cannot be rewritten such, that the elliptic integral of the second kind, ${\bf E}(z)$, does not emerge in the 
solution. The corresponding inhomogeneous solution is now obtained be Eq.~(\ref{eq:INHOM}).

We would like to end this section by a remark on simple elliptic solutions, which are sometimes also obtained in 
$x$-space. They are given by complete elliptic integrals ${\bf K}$ and ${\bf E}$ of the argument $1-x$ or $x$.
In Mellin space, they correspond to a {\it first order} factorizable problem, cf.~\cite{vonManteuffel:2017hms}
for an example. The Mellin transform 
\begin{eqnarray}
\Mvec[f(x)](N) = \int_0^1 dx x^{N-1} f(x)
\end{eqnarray}
yields
\begin{eqnarray}
\label{eq:Ksimp}
\Mvec[{\bf K}(1-z)](N)  &=& \frac{2^{4N+1}}{\displaystyle (1+2N)^2 \binom{2N}{N}^2} 
\\
\label{eq:Esimp}
\Mvec[{\bf E}(1-z)](N)  &=& \frac{2^{4N+2}}{\displaystyle (1+2N)^2 (3+2N) \binom{2N}{N}^2},
\end{eqnarray}
since
\begin{eqnarray}
{\bf K}(1-z) &=& \frac{1}{2} \frac{1}{\sqrt{1-z}} \otimes \frac{1}{\sqrt{1-z}} 
\\
{\bf E}(1-z) &=& \frac{1}{2} \frac{z}{\sqrt{1-z}} \otimes \frac{1}{\sqrt{1-z}}~. 
\end{eqnarray}
The Mellin convolution is  defined by
\begin{eqnarray}
A(x) \otimes B(x) &=& \int_0^1 dz_1 \int_0^1 dz_2 \delta(x-z_1 z_2) A(z_1) B(z_2).
\end{eqnarray}
Eqs.~(\ref{eq:Ksimp}) and (\ref{eq:Esimp}) are hypergeometric terms in $N$, which has been shown already in 
Ref.~\cite{Ablinger:2013eba} for ${\bf K}(1-x)$, see also \cite{Ablinger:2014bra}. As we outlined in 
Ref.~\cite{Ablinger:2015tua} the solution of systems of differential equations or difference equations can 
always be obtained algorithmically in the case either of those factorizes to first order. The transition to 
$x$-space is then straightforward. 
\section{Iterative non-iterative integrals}
\label{sec:3}

\vspace*{1mm} \noindent
Differential operators factorizing at first order have iterative integral solutions of the kind
\begin{eqnarray}
\label{eq:it}
F_{a_1,...,a_k}(x) = \int_0^x dy_1 f_{a_1}(y_1) \int_0^{y_1} dy_2 f_{a_2}(y_2) .... \int_0^{y_{k-1}} dy_k 
f_{a_k}(y_k),
\end{eqnarray}
where $\mathfrak{A}$ is a certain alphabet and $\forall f_l(x) \in \mathfrak{A}$. In particular, the spaces
of iterative integrals discussed in Refs.~\cite{Remiddi:1999ew,Ablinger:2011te,Moch:2001zr,Ablinger:2013eba, 
Ablinger:2014bra} are examples for this.

As well-known, the integral representation of the $_2F_1$-function in the cases having been discussed above
\begin{eqnarray}
\label{eq:hyp1}
\pFq{2}{1}{a,b}{c}{z} =  \frac{\Gamma(x)}{\Gamma(b) \Gamma(c-b)} \int_0^1 dt t^{b-1} (1-t)^{c-b-1} (1-zt)^{-a}
\end{eqnarray}
cannot be rewritten as an integral in which the $z$ dependence is just given by its boundaries.\footnote{
This will not apply to simpler cases like $\pFq{2}{1}{1,1}{2}{-z} = \ln(1+z)/z$ 
or $\pFq{2}{1}{\tfrac{1}{2},1}{\tfrac{3}{2}}{z} = \arctan(z)/z$, however.} Therfore Eq.~(\ref{eq:INHOM})
contains {\it definite} integrals, over which one integrates {\it iteratively}. We have called these {\it iterative non-iterative 
integrals} in \cite{ICMS16,Ablinger:2017bjx}. They will also occur in case the degree of non-factorization is larger than
one by virtue of the corresponding formula of the variation of the constant; the corresponding solutions of the 
homogeneous 
equations will have (multiple) integral representations with the same property like for Eq.~(\ref{eq:hyp1}).

The new iterative integrals are given by  
\begin{eqnarray}
\label{eq:ITNEW}
\mathbb{H}_{a_1,..., a_{m-1};\{ a_m; F_m(r(y_m))\},a_{m+1},...,a_q}(x) &=& \int_0^x dy_1 
\hat{f}_{a_1}(y_1) \int_0^{y_1} dy_2 ... \int_0^{y_{m-1}} dy_m \hat{f}_{a_m}(y_m) 
\nonumber\\ &&
\times F_m[r(y_m)]  H_{a_{m+1},...,a_q}(y_{m}),
\end{eqnarray}
and cases in which more than one definite integral $F_m$ appears. Here the $\hat{f}_{a_i}(y)$ are the 
usual letters of the different classes considered in 
\cite{Remiddi:1999ew,Ablinger:2011te,Ablinger:2013cf,Ablinger:2014bra} multiplied by 
hyperexponential pre-factors 
\begin{eqnarray}
r(y) y^{r_1}(1-y)^{r_2},~~~~r_i \in \mathbb{Q},~r(y) \in \mathbb{Q}[y]
\end{eqnarray}
and $F[r(y)]$ is given by
\begin{eqnarray}
F[r(y)] = \int_0^1 dz g(z,r(y)),~~~r(y) \in \mathbb{Q}[y], 
\end{eqnarray}
such that the $y$-dependence cannot be transformed into one of the integration boundaries completely.
We have chosen here $r(y)$ as a rational function because of concrete examples in this paper, which, 
however, is not necessary. 

The further analytic representation of the functions $\mathbb{H}$ will be subject to the iterated functions 
$\hat{f}_l$ and $F_m$. We will turn to this in the case of the examples (\ref{eq:INHOM}) for $\psi_{1(2)b}$ in 
Section~\ref{sec:5}.

\section{Numerical representation}
\label{sec:4}

\vspace*{1mm} \noindent
For physical applications numerical representations of the Feynman integrals have to be given. The use of 
integral-representations in {\tt Mathematica} or {\tt Maple} is possible, but usually to slow. One aims on 
efficient numerical implementations. In case of multiple polylogarithms it is available in {\tt Fortran}
\cite{Gehrmann:2001pz,Ablinger:2017tqs}, for cyclotomic polylogarithms in \cite{Ablinger:2017tqs}, where in both 
cases the method of Bernoulli-improvement is used \cite{tHooft:1978jhc}. For generalized polylogarithms a numerical 
implementation was given in \cite{Vollinga:2004sn}. All these representations are series representations. Furthermore,
there exist numerical implementations for the efficient use of harmonic sums in complex contour integral calculations
\cite{ANCONT}.

Also in case of the solutions (\ref{eq:INHOM}) analytic series representations can be given. This has been 
already the solution-strategy in \cite{Grigo:2012ji}, using power-series Ans\"atze, without further
reference to the expected mathematical structure. It turns out, that series expansions around $x = 0,1$ are 
not convergent 
in the whole interval $x \in [0,1]$. However, they have a sufficient region of overlap. Some series expansions
of the inhomogeneous solution even exhibit a singularity, cf.~\cite{Ablinger:2017bjx}, although this singularity is an 
artefact of the series expansion only. Yet these solutions can be obtained analytically and they evaluate very fast 
numerically.

The first terms of the expansion of $f_{8a}$ around $x=0$ read
\begin{eqnarray}
&& f_{8a}(x) = \nonumber\\ && 
-\sqrt{3} \Biggl[
        \pi ^3 \Biggl(
 \frac{35 x^2}{108}
-\frac{35 x^4}{486}
-\frac{35 x^6}{4374}
-\frac{35 x^8}{13122}
-\frac{70x^{10}}{59049}
-\frac{665 x^{12}}{1062882}
\Biggr)
+\Biggl(
 12 x^2
-\frac{8 x^4}{3}
\nonumber\\ && 
-\frac{8 x^6}{27}
-\frac{8 x^8}{81}
-\frac{32 x^{10}}{729}
-\frac{152 x^{12}}{6561}
\Biggr) 
{\sf Im}\Biggl[\text{Li}_3\left(\frac{e^{-\frac{i \pi}{6}}}{\sqrt{3}}\right)\Biggr]
\Biggr]
-\pi ^2 \Biggl(
1
+\frac{x^4}{9}
-\frac{4 x^6}{243}
-\frac{46 x^8}{6561}
\nonumber\\ && 
-\frac{214 x^{10}}{59049}
-\frac{5546 x^{12}}{2657205}
\Biggr)
+\Biggl(
\frac{3}{2}
+\frac{x^4}{6}
-\frac{2 x^6}{81}
-\frac{23 x^8}{2187}
-\frac{107 x^{10}}{19683}
-\frac{2773 x^{12}}{885735}
\Biggr) 
\psi ^{(1)}\hspace*{-1mm}\left(\frac{1}{3}\right)
\nonumber\\ && 
-\sqrt{3} \pi  \Biggl(
 \frac{x^2}{4}
-\frac{x^4}{18}
-\frac{x^6}{162}
-\frac{x^8}{486}
-\frac{2 x^{10}}{2187}
-\frac{19 x^{12}}{39366}
\Biggr) \ln^2(3)
-\Biggl[
        33 x^2
        -\frac{5 x^4}{4}
        -\frac{11 x^6}{54}
\nonumber\\ &&
        -\frac{19 x^8}{324}
        -\frac{751 x^{10}}{29160}
        -\frac{2227 x^{12}}{164025}
        +\pi ^2 \Biggl(
 \frac{4 x^2}{3}
-\frac{8 x^4}{27}
-\frac{8 x^6}{243}
-\frac{8 x^8}{729}
-\frac{32 x^{10}}{6561}
-\frac{152 x^{12}}{59049}
        \Biggr)
\nonumber\\ &&
        +\Biggl(
-2 x^2
+\frac{4 x^4}{9}
+\frac{4 x^6}{81}
+\frac{4 x^8}{243}
+\frac{16 x^{10}}{2187}
+\frac{76 x^{12}}{19683}
\Biggr) 
\psi^{(1)}\left(\frac{1}{3}\right)
\Biggr] \ln(x)
+\frac{135}{16}
+19 x^2
\nonumber\\ &&
-\frac{43 x^4}{48}
-\frac{89 x^6}{324}
-\frac{1493 x^8}{23328}
-\frac{132503 x^{10}}{5248800}
-\frac{2924131 x^{12}}{236196000}
-\Biggl(
        \frac{x^4}{2}-12 x^2\Biggr) \ln^2(x)
\nonumber\\ &&
-2 x^2 \ln^3(x)
+ O\left(x^{14} \ln(x)\right).
\end{eqnarray}
Likewise, one may expand around $y= 1-x = 0$ 
and obtains
\begin{eqnarray}
f_{8a}(x) &=& 
\frac{275}{12} + \frac{10}{3} y  - 25 y^2 + \frac{4}{3}  y^3  + \frac{11}{12} y^4 
+ y^5 + \frac{47}{96} y^6 + \frac{307}{960} y^7  + \frac{19541}{80640 } y^8 
\nonumber\\ &&
+ \frac{22133}{120960} y^9 
+ \frac{1107443}{7741440} y^{10} + 
 \frac{96653063}{851558400} y^{11} 
+ \frac{3127748803}{34062336000} y^{12} 
\nonumber\\ &&
+ 7 \Biggl(2 y^2 
- y^3 - \frac{1}{8} y^4  - \frac{1}{64} y^6 - \frac{1}{128} y^7  - \frac{3}{512} y^8 - 
\frac{1}{256} y^9 - \frac{47}{16384} y^{10} 
\nonumber\\ &&
- \frac{69}{32768} y^{11} - \frac{421}{262144} y^{12}
\Biggr) \zeta_3 + O(y^{13})~.
\end{eqnarray}
\begin{figure}[H]
\centering
\includegraphics[width=0.47\textwidth]{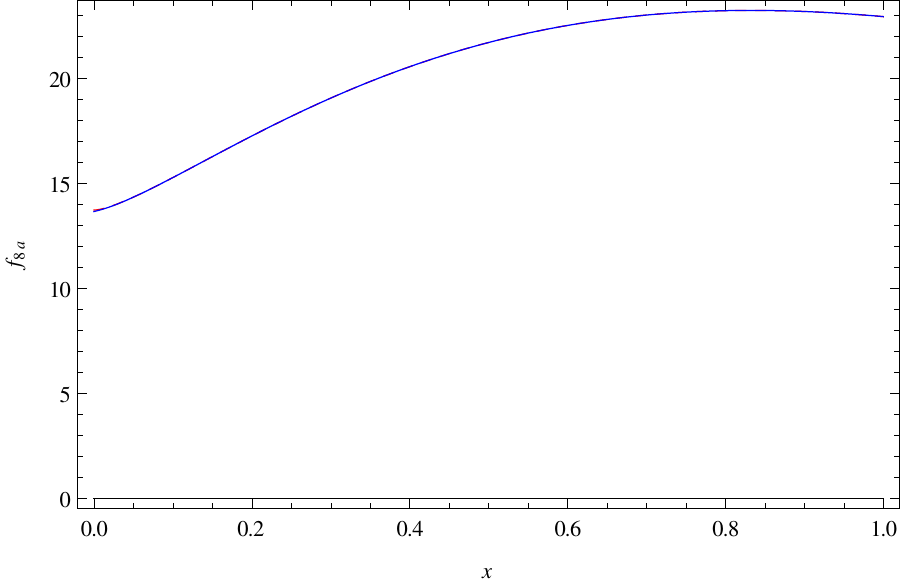}
\includegraphics[width=0.52\textwidth]{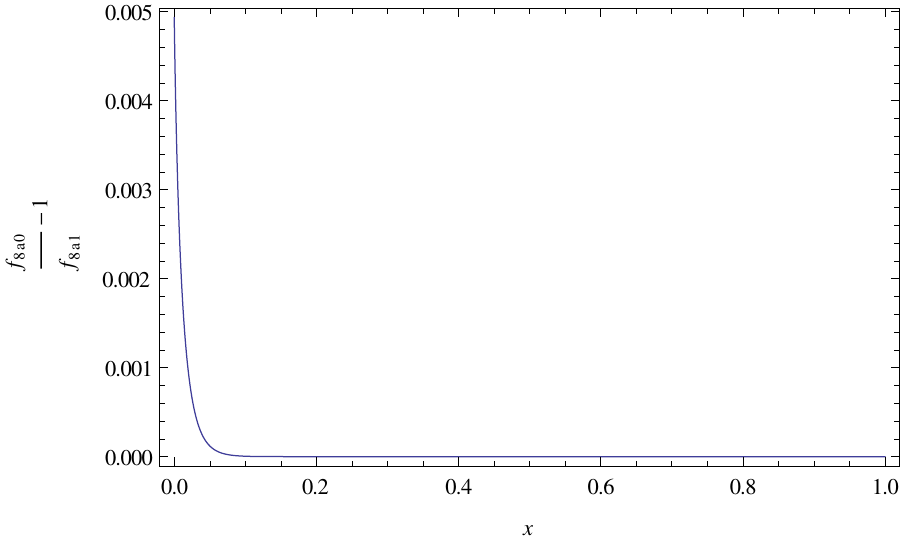}
\caption{\it \small The inhomogeneous solution of
Eq.~(\ref{eq:one}) as a function of $x$. Left panel: Red dashed line: expansion
around $x=0$; blue line: expansion around $x=1$. Right panel: illustration of the
relative accuracy and overlap of the two solutions $f_{8a}(x)$ around 0 and 1.
\label{fig:f8a}}
\end{figure}
In Figure~\ref{fig:f8a} a numerical illustration for the function $f_{8a}(x)$ is given together
with the validity of the two expansions taking into account 50 terms.
For many physics applications one would proceed in the above way and stop here. However, from the point of view 
of mathematics further interesting aspects arise to which we tun now.

\section{Representation in terms of modular forms}
\label{sec:5}

\vspace*{1mm} \noindent
The iterative non-iterative integral (\ref{eq:INHOM}) is non-iterative by virtue of the emergence of the
two complete elliptic integrals ${\bf K}(z)$ and ${\bf E}(z)$, with the modulus squared $k^2 = z(x)$ the rational 
function (\ref{eq:z1}). 
Accordingly, the second solution depends on the functions ${\bf K'}(z) = {\bf K}(1-z)$ and ${\bf E'}(z) = {\bf 
E}(1-z)$. One may re-parameterize the problem referring to the nome
\begin{eqnarray}
\label{eq:nome}
q = \exp(i\pi\tau),
\end{eqnarray}
as  the new variable with
\begin{eqnarray}
\label{qeqa}
\tau = i \frac{{\bf K}(1-z(x))}{{\bf K}(z(x))}~~~~
\text{with}~~~~\tau \in \mathbb{H} =
\left\{z \in \mathbb{C}, {\sf Im}(z) > 0\right\}.
\end{eqnarray}
All functions contributing to the solutions (\ref{eq:INHOM}, \ref{eq:ps1b}, \ref{eq:ps2b}) have now to be
translated from $x$ to $q$. 

\subsection{The mathematical framework} \label{sec:51}

\vspace*{1mm}
\noindent
For the further discussion, a series of definitions is necessary, see also 
Refs.~\cite{SERRE,COST,KF,KOECHER,KOEHLER,ONO,MILNE,RADEMACHER,DIAMOND,MART1,HIDA,IWANIEC,KILFORD,KNAPP,KNOPP,RANKIN,
SHIMURA,HECKE,SCHOENENBERG,APOSTOL,MIYAKE,OGG,ZAGIER1}.
We will use Dedekind's $\eta$-function \cite{DEDEKINDeta} 
\begin{eqnarray}
\label{eq:DEDmain}
\eta(\tau) = \frac{q^{\frac{1}{12}}}{\phi(q^2)},~~\phi(q) = \prod_{k=1}^\infty \frac{1}{1-q^k},
\end{eqnarray}
to express all quantities in the following. Here $\phi(q)$ denotes Euler's totient function 
\cite{EUL4}.
\definition
\label{def:6.1}
\noindent
Let $r = \left(r_\delta\right)_{\delta|N}$ be a finite sequence of integers indexed by the divisors 
$\delta$ of $N \in \mathbb{N} \backslash \{0\}$. The function $f_r(\tau)$ 
\begin{eqnarray}
f_r(\tau) := \prod_{\delta|N} \eta(\delta \tau)^{r_\delta},~~~\delta,N \in \mathbb{N} 
\backslash\{0\},~~r_\delta \in \mathbb{Z},
\label{eq:etaR}
\end{eqnarray}
is called {\it $\eta$-ratio}.
\rm
The $\eta$-ratios, up to differential operators in $q$, will represent all expressions in 
the following.

Let
\begin{eqnarray}
{\rm SL}_2(\mathbb{Z}) = \Biggl\{ M = \left(
\begin{array}{cc} a & b \\ c & d \end{array}
\right),~a,b,c,d \in \mathbb{Z},~\det(M) = 1 \Biggr\}.
\nonumber
\end{eqnarray}
${\rm SL}_2(\mathbb{Z})$ is the {\it modular group}.

For $g = \left(\begin{array}{cc} a & b \\ c & d \end{array} \right) \in {\rm SL}_2(\mathbb{Z})$ and $z \in \mathbb{C} \cup 
\infty$ one defines the M\"obius transformation
\begin{eqnarray}
g z \mapsto \frac{az + b}{cz + d}.
\nonumber
\end{eqnarray}
Let 
\begin{eqnarray}
S = \left(\begin{array}{rr} 0 & -1 \\ 1 & 0 \end{array}\right),~~~~\text{and}~~~~
T = \left(\begin{array}{rr} 1 &  1 \\ 0 & 1 \end{array}\right),~~~S, T \in {\rm SL}_2(\mathbb{Z}).
\nonumber
\end{eqnarray}
The polynomials of $S$ and $T$ span ${\rm SL}_2(\mathbb{Z})$.

For $N \in \mathbb{N} \backslash \{0\}$ one considers the {\it congruence 
subgroups} of 
${\rm SL}_2(\mathbb{Z})$, $\Gamma_0(N)$,
$\Gamma_1(N)$ and $\Gamma(N)$, defined by
\begin{align*}
\Gamma_0(N) &:= \left\{ \left(\begin{array}{cc} a & b \\ c & d \end{array} \right) \in {\rm SL}_2(\mathbb{Z}), 
c \equiv 0~~(\modd~N) \right\},
\\
\Gamma_1(N) &:= \left\{ \left(\begin{array}{cc} a & b \\ c & d \end{array} \right) \in {\rm SL}_2(\mathbb{Z}), 
a \equiv d \equiv 1~~(\modd~N),~~c \equiv 0~(\modd~N) \right\},
\\
\Gamma(N) &:= \left\{ \left(\begin{array}{cc} a & b \\ c & d \end{array} \right) \in {\rm SL}_2(\mathbb{Z}), 
a \equiv d \equiv 1~~(\modd~N),~~b \equiv c \equiv 0~~(\modd~N) \right\},
\end{align*}
with ${\rm SL}_2(\mathbb{Z}) 
\supseteq \Gamma_0(N) 
\supseteq \Gamma_1(N) 
\supseteq \Gamma(N)$ and $\Gamma_0(N) \subseteq \Gamma_0(M),~M|N$.

\noindent
If $N \in \mathbb{N} \backslash \{0\}$, then the {\it index} of $\Gamma_0(N)$ in 
$\Gamma_0(1)$ is
\begin{align*}
\mu_0(N) = [\Gamma_0(1) : \Gamma_0(N)] = N \prod_{p|N} \left(1 + \frac{1}{p}\right).
\nonumber
\end{align*}
The product is over the prime divisors $p$ of $N$.

{\definition
Let $x \in \mathbb{Z} \backslash \{0\}$. The analytic function $f: \mathbb{H} \rightarrow \mathbb{C}$ is a 
{\it holomorphic modular form} of {\it weight} 
{\it w} = $k$ for $\Gamma_0(N)$ and {\it character} $a \mapsto \left(\tfrac{x}{a}\right)$ 
if 
\begin{enumerate}
\item \[
f\left(\frac{az +b}{cz +d}\right) = \left(\frac{x}{a}\right) (cz+d)^k f(z),~~~\forall z \in 
\mathbb{H},~\forall \left( \begin{array}{cc} a & b \\ c &d \end{array} \right) \in \Gamma_0(N).\]
\item $f(z)$ is holomorphic in $\mathbb{H}$
\item $f(z)$ is holomorphic at the cusps of $\Gamma_0(N)$.
\end{enumerate}
Here $\left(\tfrac{x}{a}\right)$ denotes the Jacobi symbol. 
A modular form is called a {{\it cusp form}} if it vanishes at the cusps.}

For any congruence subgroup $G$ of {\rm SL}$_2(\mathbb{Z}$) a {\it cusp} of $G$ is an 
equivalence class in
$\mathbb{Q} \cup \infty$ under the action of $G$.

{\definition
A {\it meromorphic modular function} $f$ for $\Gamma_0(N)$ and weight {\it w = $k$} obeys
\begin{enumerate}
\item $f(\gamma z) = (cz+d)^{k} f(z),~~~\forall z \in \mathbb{H}~~\text{and}~~\forall \gamma \in \Gamma_0(N)$ 
\item $f$ is meromorphic in $\mathbb{H}$
\item $f$ is meromorphic at the cusps of $\Gamma_0(N)$.
\end{enumerate}
The $q$-expansion of a meromorphic modular form has the form
\begin{equation}
f^*(q) = \sum_{k=-N_0}^\infty a_k q^k,~~~\text{for~some}~~N_0 \in \mathbb{N}.
\nonumber
\end{equation}}

\noindent
{\lemma
The set of functions ${\cal M}(k; N; x)$ for $\Gamma_0(N)$ and character $x$, defined above, 
forms a finite dimensional vector space over $\mathbb{C}$. In particular, for
any non-zero function $f\in {\cal M}(k; N; x)$ we have
\begin{eqnarray}
{\rm ord}(f) \leq b = \frac{k}{12} \mu_0(N),
\nonumber
\label{eq:DIMV}
\end{eqnarray}
cf. e.g.~{\rm \cite{OGG,KOECHER,KOEHLER}}}. The bound 
(\ref{eq:DIMV}) on the dimension can be refined, see
e.g.~\cite{HECKE,MILNE,SCHOENENBERG,MIYAKE} for details.\footnote{
The dimension of the corresponding vector space can be also calculated using the {\tt Sage} program
by W.~Stein \cite{STEIN}.}.
The number of independent modular forms $f \in {\cal M}(k; N; x)$ is $\leq b$, allowing for a basis 
representation in finite terms.

For any $\eta$-ratio $f_r$ one can prove that there exists a minimal integer $l \in \mathbb{N}$,
an integer $N \in \mathbb{N}$ and a character $x$ such that
{
\begin{eqnarray}
\bar{f}_r(\tau) = \eta^l(\tau) f_r(\tau) \in {\cal M}(k; N; x)
\nonumber
\end{eqnarray}}
is a {holomorphic modular form.} All quantities which are expanded in $q$-series below will 
be first brought 
into 
the 
above 
form. In some cases one has $l =0$. This form is of importance to obtain
Lambert-Eisenstein series \cite{LAMBERT,EISENSTEIN}, 
which can be rewritten in terms of elliptic polylogarithms \cite{ELLPOL}.

A basis of the vector space of holomorphic modular forms is given by 
the associated Lambert-Eisenstein series with character and binary products
thereof \cite{OGG,COST}.

The Lambert-Eisenstein series are given by
\begin{eqnarray}
\label{eq:LAM1}
\sum_{k=1}^\infty \frac{k^\alpha q^k}{1 - q^k} = \sum_{k=1}^\infty \sigma_{\alpha}(k) 
q^k,~~~\sigma_\alpha(k) = 
\sum_{d|k} d^\alpha,~~~\alpha \in \mathbb{N}.
\end{eqnarray} 
They can be rewritten in terms of elliptic polylogarithms, which we will use rather as a 
frame in the following, 
\begin{eqnarray}
\label{eq:ELP1}
{\rm 
ELi}_{n;m}(x;y;q) 
:=
\sum_{k = 
1}^\infty
\sum_{l = 
1}^\infty 
\frac{x^k}{k^n} 
\frac{y^l}{l^m} 
q^{kl}
\end{eqnarray}
by
\begin{eqnarray}
\sum_{k=1}^\infty \frac{k^\alpha q^k}{1 - q^k} = \sum_{k=1}^\infty k^\alpha \Li_0(q^k) = \sum_{k,l = 
1}^\infty k^\alpha q^{kl} = {\rm ELi}_{-\alpha;0}(1;1;q),
\end{eqnarray} 

\noindent with $\Li_0(x) = x/(1-x)$. It also appears useful to define \cite{Adams:2016xah},
\begin{eqnarray}
\label{eq:ELP3}
\overline{E}_{n;m}(x;y;q) = \Biggl\{
\renewcommand{\arraystretch}{1.5}
\begin{array}{ll}
\tfrac{1}{i} [\ELI_{n;m}(x;y;q) - \ELI_{n;m}(x^{-1};y^{-1};q)], & n+m~~\text{even}\\
~~\ELI_{n;m}(x;y;q) + \ELI_{n;m}(x^{-1};y^{-1};q), & n+m~~\text{odd}.
\\
\end{array} 
\renewcommand{\arraystretch}{1}
\end{eqnarray}

\noindent
The multiplication relation of elliptic polylogarithms is given by \cite{ELLPOL}
\begin{align}
\label{eq:ELP4a}
&{\ELI}_{n_1,...,n_l;m_1,...,m_l;0,2o_2,...,2o_{l-1}}(x_1,...,x_l;y_1,...,y_l;q) = 
\nonumber\\
& 
{\ELI}_{n_1;m_1}(x_1;y_1;q)
{\ELI}_{n_2,...,n_l;m_2,...,m_l;2o_2,...,2o_{l-1}}(x_2,...,x_l;y_2,...,y_l;q), 
\end{align}
with
\begin{eqnarray}
\label{eq:ELP2}
&&\ELI_{n_,...,n_l;m_1,...,m_l;2o_1,...,2o_{l-1}}(x_1,...,x_l;y_1,...y_l;q)
\\ 
&&= \sum_{j_1=1}^\infty ...
\sum_{j_l=1}^\infty 
\sum_{k_1=1}^\infty ...
\sum_{k_l=1}^\infty 
\frac{x_1^{j_1}}{j_1^{n_1}} ...
\frac{x_l^{j_l}}{j_l^{n_l}} 
\frac{y_1^{k_1}}{k_1^{m_1}} 
\frac{y_l^{k_l}}{k_l^{m_l}} 
\frac{q^{j_1 k_1+...+q_l k_l}}{\prod_{i=1}^{l-1}(j_ik_i+...+j_lk_l)^{o_{i}}},l > 0.
\nonumber
\end{eqnarray}

The logarithmic integral of an elliptic polylogarithm is given by
\begin{align}
\label{eq:ELP5}
&\ELI_{n_1,...,n_l;m_1,...,m_l;2(o_1+1),2o_2,...,2o_{l-1}}(x_1,...,x_l;y_1,...,y_l;q) = 
\nonumber\\
& 
\hspace*{2cm}
\int_0^q 
\frac{dq'}{q'} 
\ELI_{n_1,...,n_l;m_1,...,m_l;2o_1,...,2o_{l-1}}(x_1,...,x_l;y_1,...,y_l;q'). 
\end{align}
Similarly, cf.~\cite{Adams:2016xah},
\begin{align}
&\overline{E}_{n_1,...,n_l;m_1,...,m_l;0,2o_2,...,2o_{l-1}}(x_1,...,x_l;y_1,...,y_l;q) =
\nonumber\\
& \hspace*{2cm}
\overline{E}_{n_1;m_1}(x_1;y_1;q)
\overline{E}_{n_2,...,n_l;m_2,...,m_l;2o_2,...,2o_{l-1}}(x_1,...,x_l;y_1,...,y_l;q)
\\
&\overline{E}_{n_1,...,n_l;m_1,...,m_l;2(o_1+1),2o_2,...,2o_{l-1}}(x_1,...,x_l;y_1,...,y_l;q) =
\nonumber\\
& \hspace*{3cm}
\int_0^q \frac{dq'}{q'}\overline{E}_{n_1,...,n_l;m_1,...,m_l;2o_1,...,2o_{l-1}}(x_1,...,x_l;y_1,...,y_l;q') 
\end{align}
holds.

The integral over the product of two more general elliptic polylogarithms is given by
\begin{eqnarray}
\int_0^q \frac{d\bar{q}}{\bar{q}} \ELI_{m,n}(x,\bar{q}^a,\bar{q}^b) 
\ELI_{m',n'}(x',\bar{q}^{a'},\bar{q}^{b'})
&=& 
\sum_{k=1}^\infty
\sum_{l=1}^\infty
\sum_{k'=1}^\infty
\sum_{l'=1}^\infty
\frac{x^k}{k^m}
\frac{x'^k}{k'^{m'}}
\frac{{q}^{al}}{l^{n}}
\frac{{q}^{a'l'}}{l'^{n}}
\nonumber\\ && \times
\frac{{q}^{bkl + b'k'l'}}{al + a'l' + bkl +bk'l'}.
\end{eqnarray}
Integrals over other products are obtained accordingly.

\noindent
In the derivation often the argument $q^m,~~m \in \mathbb{N}, m > 0$, appears, which 
shall be mapped to the variable $q$. We do this for the Lambert series using the
replacement
\begin{eqnarray}
\label{eq:xm}
\Li_0(x^m) =
\frac{x^m}{1-x^m} = \frac{1}{m} \sum_{k=1}^m \frac{\rho_m^k x}{1 - \rho_m^k x} = \frac{1}{m} \sum_{k=1}^m \Li_0(\rho_m^k 
x),
\end{eqnarray} 
with 
\begin{eqnarray}
\rho_m = \exp\left(\frac{2\pi i}{m}\right).
\end{eqnarray} 
One has
\begin{eqnarray}
\label{eq:CYC}
\sum_{k=1}^\infty \frac{k^\alpha q^{mk}}{1 - q^{mk}} = \ELI_{-\alpha;0}(1;1;q^m) =
\frac{1}{m^{\alpha+1}} \sum_{n=1}^m 
{\rm ELi}_{-\alpha;0}(\rho_m^n;1;q)~.
\end{eqnarray}
Relations like (\ref{eq:xm}, \ref{eq:CYC}) and similar ones are the sources of the $m$th roots 
of unity, which correspondingly appear in the parameters of the elliptic polylogarithms  through the 
Lambert series. 

Furthermore, the following sums occur
\begin{eqnarray}
\label{eq:extELP1}
\sum_{m=1}^\infty \frac{(am+b)^l q^{am+b}}{1 -  q^{am+b}} &=& \sum_{n=1}^l \binom{l}{n} a^{n} b^{l-n}
\sum_{m=1}^\infty \frac{m^n q^{am+b}}{1-q^{am+b}},~~~a, l \in \mathbb{N},~~~b \in \mathbb{Z}
\end{eqnarray}
and 
\begin{eqnarray}
\label{eq:extELP2}
\sum_{m=1}^\infty \frac{m^n q^{am+b}}{1-q^{am+b}} = \ELI_{-n;0}(1;q^b;q^a) = \frac{1}{a^{n+1}} \sum_{\nu = 
1}^a \ELI_{-n;0}(\rho_a^\nu;q^b;q)~.
\end{eqnarray}

Likewise, one has
\begin{eqnarray}
\label{eq:extELP2a}
\sum_{m=1}^\infty \frac{(-1)^m m^n q^{am+b}}{1-q^{am+b}} &=& 
\ELI_{-n;0}(-1; q^b; q^a) 
\\ &=&
\frac{1}{a^{n+1}} \left\{
\sum_{\nu = 1}^{2a} \ELI_{-n;0}(\rho_{2a}^\nu;q^b;q)
-\sum_{\nu = 1}^{a} \ELI_{-n;0}(\rho_{a}^\nu;q^b;q)\right\}.
\nonumber
\end{eqnarray}
In intermediate representations also Jacobi symbols appear, obeying the identities
\begin{eqnarray}
\label{eq:JACs}
\left(\frac{-1}{(2k)\cdot n + (2l+1)}\right) = (-1)^{k+l};~~~~
\left(\frac{-1}{a b}\right) =
\left(\frac{-1}{a}\right)
\left(\frac{-1}{b}\right).
\end{eqnarray}
In the case of an even value of the denominator one may factor $\left(\tfrac{-1}{2}\right) = 1$ and 
consider the case of the remaining odd-valued denominator.  

We found also Lambert series of the kind 
\begin{eqnarray}
\label{eq:extELP2b}
\sum_{m=1}^\infty \frac{q^{(c-a)m}}{1-q^{cm}} &=& \ELI_{0;0}(1;q^{-a};q^c) = \frac{1}{c} \sum_{n=1}^c
\ELI_{0;0}(\rho_c^n;q^{-a};q)  
\\
\sum_{m=1}^\infty (-1)^m \frac{q^{(c-a)m}}{1-q^{cm}} &=& \ELI_{0;0}(1;-q^{-a};q^c) = \frac{1}{c} \sum_{n=1}^c
\ELI_{0;0}(\rho_c^n;-q^{-a};q),
\nonumber\\ &&
~~~a,c \in \mathbb{N} \backslash \{0\}  
\end{eqnarray}
in intermediate steps of the calculation.

Also the functions
\begin{eqnarray}
\label{eq:Y}
Y_{m,n,l} &:=& \sum_{k=0}^\infty \frac{(mk+n)^{l-1} q^{mk+n}}{1-q^{mk+n}} 
\nonumber\\ &=&
n^{l-1} \Li_0(q^n) + \sum_{j=0}^{l-1} \binom{l-1}{j} n^{l-1-j} m^j \ELI_{-j;0}(1;q^n;q^m)
\\
\label{eq:Z}
Z_{m,n,l} &:=& \sum_{k=1}^\infty \frac{k^{m-1} q^{nk}}{1-q^{lk}}  = \ELI_{0;-(m-1)}(1;q^{n-l};q^l)
\\
T_{m,n,l,a,b} &:=& \sum_{k=0}^\infty \frac{(mk + n)^{l - 1} q^{a (mk + n)}}{1 - q^{b (mk + n)}} 
= 
n^{l-1} q^{n(a-b)} 
\Li_0\left(q^{nb}\right) 
\nonumber\\ &&
+ q^{n(a-b)} \sum_{j=0}^{l-1} 
\binom{l-1}{j} m^j n^{l-1-j} \ELI_{-j;0}\left(q^{m(a-b)};q^{nb};q^{mb}\right)
\label{eq:T}
\end{eqnarray}
contribute. Note that (part of) the parameters $(x;y)$ of the elliptic polylogarithms 
can become $q$-dependent, unlike the case in \cite{BLOCH2,Adams:2016xah}. 
The elliptic polylogarithms rather form a suitable frame here, while we give preference to
the Lambert-Eisenstein series. The $q$-dependence of $x (y)$ does not spoil the integration
relations, which can be generalized in case factors $1/\eta(\tau)$ do not occur in addition.

\subsection{The \boldmath $q$-representation of the inhomogeneous solution} \label{sec:52}

\vspace*{1mm}
\noindent
Now we turn to (\ref{eq:INHOM}) again and express all quantities in terms of the variable 
$q$.

The modulus is given by 
\begin{eqnarray}
k  &=& \frac{4 \eta^{8}(2\tau) \eta^4\left(\frac{\tau}{2}\right)}{\eta^{12}(\tau)},~~~~~~
k' = \frac{\eta^{4}(2\tau) \eta^{8}\left(\frac{\tau}{2}\right)}{\eta^{12}(\tau)},
\end{eqnarray}
which implies the following relation by $k' = \sqrt{1-k^2}$ for $\eta$ functions
\begin{eqnarray}
\label{eq:ID4}
1 = \frac{\eta^8\left(\frac{\tau}{2}\right) \eta^8(2\tau)}{\eta^{24}(\tau)} \left[16 \eta^8(2\tau)
+ \eta^8\left(\frac{\tau}{2}\right) \right]~.
\end{eqnarray}
The elliptic integral of the first kind has the representation \cite{TRICOMI}, sometimes
also written using Jacobi's $\vartheta_i$-functions \cite{JAC2},
\begin{eqnarray}
\label{eq:KKp}
{\bf K}(k^2) &=& \frac{\pi}{2} 
\frac{\eta^{10}(\tau)}{\eta^4\left(\frac{1}{2}\tau\right) \eta^4(2 \tau)},~~~~~
{\bf K'}(k^2) = - \frac{1}{\pi} {\bf K}(k^2)~ \ln(q)~.
\end{eqnarray}

The elliptic integrals of the 2nd kind, ${\bf E}$ and ${\bf E}'$ are given by
\cite{ERDELYI2,ABRSTE}
\begin{eqnarray}
{\bf E}(k^2)   &=& {\bf K}(k^2) + \frac{\pi^2 q}{{\bf K}(k^2)} \frac{d}{dq} \ln\left[\vartheta_4(q)\right] 
\label{eq:EMain}
\end{eqnarray}
and the Legendre identity \cite{LEGENDRE}
\begin{eqnarray}
\label{eq:LEGEND}
{\bf K}(z) {\bf E}(1-z) + {\bf E}(z) {\bf K}(1-z) - {\bf K}(z) {\bf K}(1-z) = \frac{\pi}{2},
\end{eqnarray}
to express {\bf E}$'$,
\begin{eqnarray}
{\bf E}'(k^2)   &=& \frac{\pi}{2{\bf K}(k^2)} \left[1 + 2 \ln(q)~q\frac{d}{dq} 
\ln\left[\vartheta_4(q)\right]\right],
\label{eq:EMainPR}
\end{eqnarray}
where the Jacobi $\vartheta$ functions are given by 
\begin{eqnarray}
\vartheta_2(q) &=& \frac{2 \eta^2(2\tau)}{\eta(\tau)},~~~
\vartheta_3(q) = \frac{\eta^5(\tau)}{\eta^2\left(\frac{1}{2}\tau\right) \eta^2(2 \tau)},
~~~\vartheta_4(q) = \frac{\eta^2\left(\frac{\tau}{2}\right)}{\eta(\tau)}.
\end{eqnarray}

We have now to determine the kinematic variable $x = x(q)$ analytically. This is not always possible 
for other choices of the definition of $q$, cf.~\cite{Adams:2014vja}. In the present case, however,
a cubic Legendre-Jacobi transformation \cite{LEGENDRE1,JAC5}\footnote{This is, besides the 
well-know   Landen transformation \cite{TRICOMI,LANDENGAUSS}, the next higher modular 
transformation; for a survey cf.~\cite{CAYLEY}. Also for the 
hypergeometric function $\pFq{2}{1}{\tfrac{1}{r},1-\tfrac{1}{r}}{1}{z(x)}$ there are rational 
modular transformations \cite{MAIER}.} allows the solution.
Following \cite{BORWEIN1,Bailey:2008ib,Broadhurst:2008mx,JOYCE}
\begin{eqnarray}
\label{eq:cubic1}
\frac{16 y}{(1-y) (1+3y)^3} = \frac{\vartheta_2^4(q)}{\vartheta_3^4(q)}
\end{eqnarray}
is solved by
\begin{eqnarray}
\label{eq:cubic2}
y = \frac{\vartheta_2^2(q^3)}{\vartheta_2^2(q)} \equiv - \frac{1}{3\overline{x}} = \frac{1}{3x}.
\end{eqnarray}
Both the expressions (\ref{eq:cubic1}, \ref{eq:cubic2}) are modular functions. For definiteness,
we consider the range in $q$
\begin{eqnarray}
\label{eq:kincond}
q \in [-1,1]~~~\text{which~corresponds~to}~~~
y \in \left[0, \tfrac{1}{3}\right],~~~~x \in \left[1, +\infty\right[
\end{eqnarray}
in the following. Here the variable $x$ lies in the unphysical region. However, the nome $q$ has to obey the
condition (\ref{eq:kincond}). Other kinematic regions can be reached performing analytic
continuations.

One obtains 
\begin{eqnarray}
x = \frac{1}{3} \frac{\eta^{4}(2\tau)
\eta^{2}(3\tau)}{\eta^2(\tau) \eta^4(6\tau)}.
\end{eqnarray}
By this all ingredients of the inhomogeneous solution (\ref{eq:INHOM}) can now be rewritten in $q$.
Using the on-line encyclopedia of integer sequences
\cite{OEIS} one finds in particular for entry {\tt A256637}
\begin{eqnarray}
\sqrt{(1-3x)(1+x)}     &=& \frac{1}{i\sqrt{3}} 
\left. 
\frac{
\eta\left(\tfrac{\tau}{2}\right)
\eta\left(\tfrac{3\tau}{2}\right) \eta(2\tau) \eta(3\tau)} 
{\eta(\tau)\eta^3(6\tau)}\right|_{q\rightarrow -q}
\end{eqnarray}
and for terms in the inhomogeneity and the Wronskian {\tt  A187100, A187153} \cite{OEIS} 
\begin{eqnarray}
\frac{1}{1-x}    &=&  -3 \frac{\eta^2(\tau) \eta\left(\tfrac{3}{2}\tau\right) 
\eta^3(6\tau)}{\eta^3\left(\tfrac{1}{2}\tau\right) \eta(2\tau) \eta^2(3\tau)}
\\
\frac{1}{1 - 3x} &=&  -\frac{\left[\eta(\tau) \eta\left(\tfrac{3}{2}\tau\right) 
\eta^2(6\tau)\right]^3}{\eta\left(\tfrac{1}{2}\tau\right) \eta^2(2\tau) \eta^9(3\tau)}.
\end{eqnarray}
This method can be applied since the $q$-series of the associated holomorphic 
modular form to these expressions factoring off a power of $1/\eta(\tau)$
is determined by a finite number of expansion coefficients.

Next we would like to investigate which kind of modular form the solution $\psi(x)$ is. Some of its building
blocks, like ${\bf K}$, are holomorphic modular forms  \cite{SERRE,COST}, while others, like  ${\bf E}$, are 
meromorphic
modular forms. In case a solution can be thoroughly expressed by holomorphic modular forms, as e.g. 
in case of the sun-rise graph studied in Refs.~\cite{BLOCH2,Adams:2014vja}, one has then the possibility
to express the result in terms of polynomials of Lambert--Eisenstein series \cite{LAMBERT,EISENSTEIN}, which
are given by elliptic polylogarithms \cite{ELLPOL} and their generalizations, 
cf. e.g.~\cite{Adams:2016xah} an references therein.

The elliptic integral of the first kind can be expressed by $E$ or $\overline{E}$-functions only.
\begin{eqnarray}
\label{eq:Kmod}
{\bf K}(z) &=& \frac{\pi}{2} \left[1 + 2 \overline{E}_{0;0}(i;1;q)\right],
\end{eqnarray}
On the other hand, this is not the case for $1/{\bf K}(z)$, a function needed to represent
${\bf E}$:
\begin{eqnarray}
\frac{1}{{\bf K}(z)} &=& \frac{2}{\pi \eta^{12}(\tau)}
\Biggl\{
\frac{5}{48} \Biggl\{
1-24 \ELI_{0;-1}(1;1;q)
-4
\Bigl[1
-\frac{3}{2}
\Bigl[\ELI_{0;-1}(1;1;q)
\nonumber\\ &&
+\ELI_{0;-1}(1;i;q)
+\ELI_{0;-1}(1;-1;q)+\ELI_{0;-1}(1;-i;q)\Bigr]
\Bigr]\Biggr\}
\Biggl\{-1
\nonumber\\ &&
+4
\Bigl[
-\frac{1}{2}\Bigl[\ELI_{-2;0}(i;1/q;q)
+\ELI_{-2,0}(-i;1/q;q)\Bigr]+
\Bigl[\ELI_{-1;0}(i;1/q;q)
\nonumber\\ &&
+\ELI_{-1;0}(-i;1/q;q)\Bigr]
-\frac{1}{2} \Bigl[\ELI_{0,0}(i;1/q;q)
+\ELI_{0,0}(-i;1/q;q)\Bigr]
\Bigr]\Biggr\}
\nonumber\\ &&
-\frac{1}{16}\Biggl\{5
+4\Bigl[-\frac{1}{2}
  \Bigl[\ELI_{-4;0}(i;1/q;q)+\ELI_{-4;0}(-i;1/q;q)\Bigr]
\nonumber\\ &&
+2\Bigl[\ELI_{-3;0}(i;1/q;q)+\ELI_{-3,0}(-i;1/q;q)\Bigr]
-3\Bigl[\ELI_{-2;0}(i;1/q;q)
\nonumber\\ &&
+\ELI_{-2,0}(-i;1/q;q)\Bigr]
+2\Bigl[\ELI_{-1;0}(i;1/q;q)+\ELI_{-1;0}(-i;1/q;q)\Bigr]
\nonumber\\ &&
-\frac{1}{2}\Bigl[\ELI_{0;0}(i;1/q;q)+\ELI_{0,0}(-i;1/q;q)\Bigr]
\Biggr\}\Biggr\}.
\label{eq:ONEoK}
\end{eqnarray}
Here and in a series of other building blocks the factor $1/\eta^{12}(\tau)$ emerges
through which the corresponding quantity becomes a meromorphic modular form \cite{Ablinger:2017bjx}.

Still one has to express the inhomogeneities of the corresponding differential equations.
They are given by harmonic polylogarithms $H_{\vec{a}}(x)$ and rational pre-factors in $x$.
In the variable $q=q(x)$ they will be different, cf.~\cite{Adams:2014vja,Ablinger:2017bjx}, depending on the 
definition of $q$. 

Since for the $q$-series of $1/\eta(\tau)$ no closed form expression of the expansion coefficients
is known, one cannot write down a closed form integration relation for polynomials out of 
quantities like this, unlike the case for polynomials out of Lambert-Eisenstein series, see 
Ref.~\cite{Ablinger:2017bjx} for details.
Therefore, a closed analytic solution of the inhomogeneous solution using structures like
elliptic polylogarithms, cf.~Section~\ref{sec:51}, cannot be given. Yet, one may use $q$-series
in the numerical representation expanding to a certain power. This, however, is equivalent to the 
numerical representation given in Section~\ref{sec:4}, where no further analytic continuation is necessary.
\section{The \boldmath $\rho$-parameter}\label{sec:6}

\vspace*{1mm}
\noindent 
Finally we would like to present numerical results on the $\rho$-parameter with a finite 
quark mass ratio, given in Ref.~\cite{Blumlein:2018aeq}.
The $\rho$-parameter is defined by
\begin{eqnarray}
\rho = 1 + 
 \frac{\Pi_T^Z(0)}{M_Z^2}
-\frac{\Pi_T^W(0)}{M_W^2} \equiv 1 + \Delta \rho,
\end{eqnarray}
with $\Pi_T^k(0)$ the respective transversal self energies at zero momentum and $M_k$ the 
masses of the $Z$ and $W$ bosons. Here the correction is given by
\begin{equation}
\Delta \rho = \frac{3 G_F m_t^2}{8 \pi^2 \sqrt{2}} \biggl(
\delta^{(0)}+\frac{\alpha_s}{\pi} \delta^{(1)}
+\left(\frac{\alpha_s}{\pi}\right)^2 \delta^{(2)}
+{\cal O}(\alpha_s^3)\biggr),
\label{eq:deltarho}
\end{equation}
where $G_F$ is the Fermi-constant, $m_t$ denotes the heavy fermion mass, and $x = 
m_b^2/m_t^2$ the ratio of the masses of the light and the heavy partner squared. 

The radiative corrections allow to set limits on 
heavy fermions in case of doublet mass splitting, which was important to
determine the precise mass region of the top-quark \cite{Veltman:1977kh}.
Radiative corrections were calculated in  
Refs.~\cite{Ross:1975fq,Veltman:1977kh,Avdeev:1994db,Chetyrkin:1995ix,Boughezal:2004ef,
Chetyrkin:2006bj,Boughezal:2006xk,Grigo:2012ji}. In Ref.~\cite{Ablinger:2017bjx}
we calculated the analytic form of the yet missing master integrals. They can now be 
evaluated numerically starting from a complete analytic representation. We insert our 
results into the representation given in \cite{Grigo:2012ji}.

\begin{figure}[H]
\begin{center}
\includegraphics[width=0.6\textwidth]{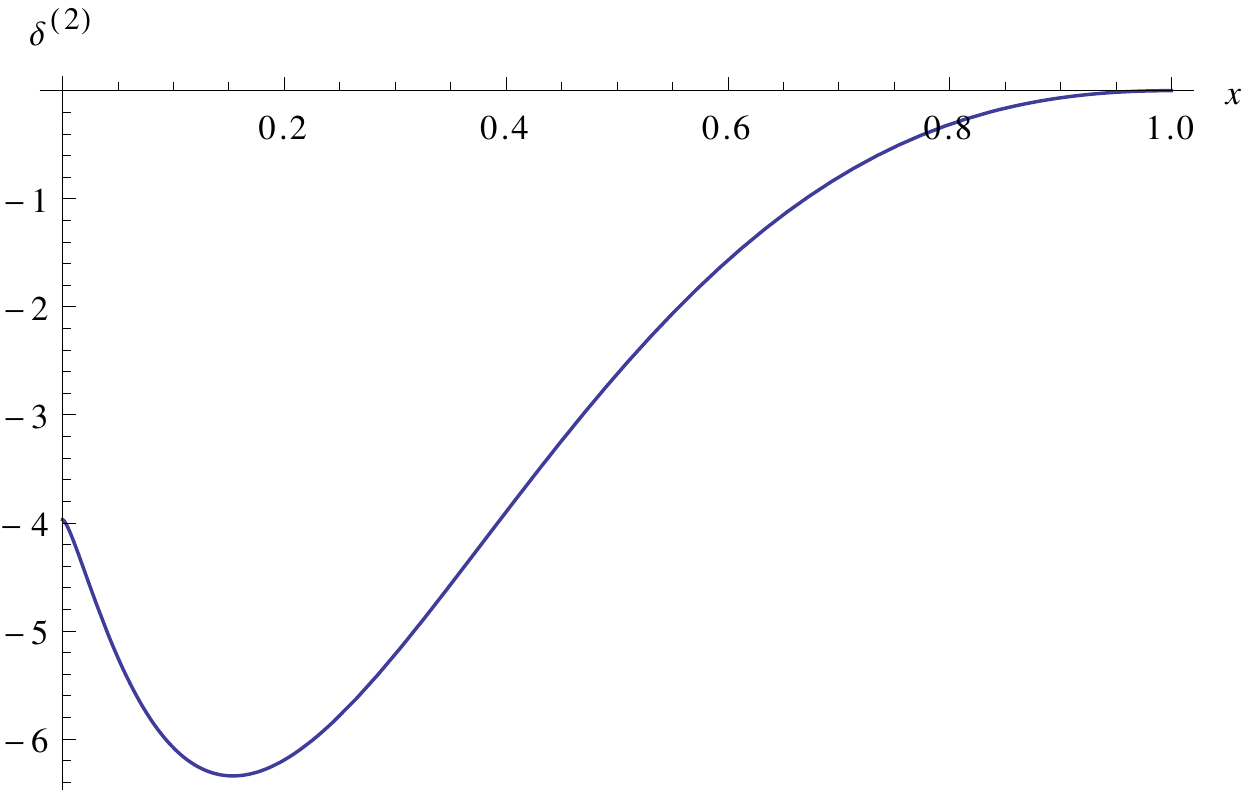}
\end{center}
\caption[]{\it The two-mass contributions to $\delta^{(2)}$ as a function of $x$.}
\label{fig:delta2}
\end{figure}

The expression for the $\delta^{(2)}$, 
Eq.~(\ref{eq:deltarho}),
in terms of the master integrals in the $\overline{\rm MS}$ scheme, is given by Eq.~(\ref{eq:d2}), where
we only show the contributions due to the  iterative non-iterative integrals.
\begin{eqnarray}
\label{eq:d2}
\delta^{(2)}(x) &=& \dots + C_F \left(C_F-\frac{C_A}{2}\right) \Biggl[
\frac{11-x^2}{12 (1-x^2)^2} f_{8a}(x)
+\frac{9-x^2}{3 (1-x^2)^2} f_{9a}(x)
\nonumber\\ &&
+\frac{1}{12} f_{10a}(x)
+\frac{5-39 x^2}{36 (1-x^2)^2} f_{8b}(x)
+\frac{1-9 x^2}{9 (1-x^2)^2} f_{9b}(x)
+\frac{x^2}{12} f_{10b}(x)
\Biggr]
\nonumber\\ &&
+\frac{C_F T_F}{9 (1-x^2)^3} \Biggl[
(5 x^4-28 x^2-9) f_{8a}(x)
+\frac{1-3 x^2}{3 x^2} (9 x^4+9 x^2-2) f_{8b}(x)
\nonumber\\ &&
+(9-x^2) (x^4-6 x^2-3) f_{9a}(x)
+\frac{1-9 x^2}{3 x^2} (3 x^4+6 x^2-1) f_{9b}(x)
\Biggr].
\end{eqnarray}
The different functions $f_i(a)$ are given in Ref.~\cite{Ablinger:2017bjx}.
The behaviour of the correction term $\delta^{(2)}(x)$ is shown in Figure~\ref{fig:delta2}.
The color factor signals that it stems from the non-planar part of the problem.
In the limit of $m_t \rightarrow \infty$ the numerical value 
$\delta^{(2)}(0) = -3.969$ is obtained in agreement with \cite{Avdeev:1994db}.
In the limit of zero mass splitting the correction vanishes.

\section{Conclusions}\label{sec:7}

\vspace*{1mm}
\noindent 
In the analytic calculation of zero- and single-scale Feynman diagrams in the most simple cases
iterative integral and indefinite nested sum representations are sufficient. Here either the system
of differential or difference equations factorizes to first order \cite{Blumlein:2018cms}. All these 
cases can be solved algorithmically, cf.~\cite{Ablinger:2015tua}, in whatsoever basis. The function spaces, which 
represent the solutions for the cases having been studied so far, are completely known and the 
associated numerical implementations are widely available.

At present an important target of research are the cases in which the level of non-factorization
is of second or higher order. Also in these cases the general structure of the formal solutions is 
known. In case of the differential equations they are given by the variation of constant, over the 
solutions of the homogeneous equations. Here the latter ones have no iterative solutions. They can be
written as (multiple) Mellin-Barnes \cite{MB} integrals \cite{Blumlein:2010zv} and by this cast into a multiple
integral representation in which the next integration variable cannot be completely transformed into
the integral boundaries. Therefore, these integrals are of non-iterative character. In summary, one 
obtains iterative integrals over these non-iterative integrals as the main structure 
\cite{ICMS16,Ablinger:2017bjx}.

From the mathematical point of view one would like to understand the non-iterative integrals emerging
on the different levels of non-factorization in more detail. In the 2nd order case the corresponding differential 
equations have $_2F_1$-solutions with specific rational parameters and rational functions in $x$ as argument.
This is generally due to the fact that the corresponding differential equations have more than three singularities.
There is a decision algorithm, cf.~\cite{IVH,VANH1,Ablinger:2017bjx}, whether or not the $_2F_1$-solutions
can be mapped on complete elliptic integrals or not. Furthermore, one may investigate using the criteria given in 
\cite{RHEUN1,RHEUN2} whether representations in terms of complete elliptic integrals of the first kind are sufficient
in special cases. In the elliptic case one may consider representation in terms of modular forms, which are in 
general meromorphic. A sub-class of only holomorphic modular forms, cf. e.g. \cite{BLOCH2,Adams:2016xah}, also exists 
in a series of interesting cases. Finally, complete elliptic integrals of the first and second kind with argument $x$ 
or $(1-x)$ do not form a 2nd order problem, if considered in $N$ space, where they have a representation in 
hypergeometric terms.

The level of non-factorization for single-scale Feynman integrals at second order is widely understood and 
throughly tied up with $_2F_1$-solutions. Their properties allow to derive also analytic solutions. Corresponding 
series expansions in the complex plane allow for numerical implementations since their convergence regions 
do sufficiently overlap. 

Much less is known in case of third and higher order non-factorization. Cases of this kind will emerge in 
future calculations. Here one is not advised to apply the pure {\it integral} approach of differential equations 
\cite{DEQ,Ablinger:2015tua}. To recognize the nature of the integrals contributing here it is useful to apply
the dispersive approach to the corresponding integrals first \cite{DISP}. Even multiple cuts may be necessary
to unravel the emerging structures. In this way, once again, {\it non-iterative} integrals are obtained. This
has been the easiest approach to solve the sun-rise graph also, cf.~\cite{Laporta:2004rb}. This method will 
be of use to unravel further levels and to establish the links needed to known mathematical structures or at 
least to guide the way to work out the corresponding mathematics, if it is not  know yet.

Again the analytic calculation of Feynman integrals shows the rich mathematical structures behind these 
quantities and leads to an intense cooperation between theoretical physics, different branches of mathematics
and computer algebra. During the last 30 years an enormous development has been taking place, but much more is going 
to come.

\vspace*{3mm} \noindent
{\bf Acknowledgment.}~I would like to thank J.~Ablinger, A.~De Freitas, M.~van Hoeij, E.~Imamoglu, 
P.~Marquard, C.G.~Raab, C.-S.~Radu, and C.~Schneider for collaboration in two projects and 
A.~Behring, D.~Broadhurst, H.~Cohen, G.~K\"ohler, P.~Paule, E.~Remiddi,  
M.~Steinhauser, J.-A.~Weil, 
S.~Weinzierl and D.~Zagier for discussions.

\end{document}